\documentclass[conference,10pt]{IEEEtran}
\IEEEoverridecommandlockouts
\usepackage{cite}
\usepackage{amsmath,amssymb,amsfonts}
\usepackage{algorithmic}
\usepackage{graphicx}
\usepackage{textcomp}
\usepackage{float}
\usepackage{xcolor}
\usepackage{threeparttable}
\usepackage{enumitem}
\usepackage[mathscr]{euscript}
\usepackage{mathtools}
\usepackage{setspace}
\usepackage{array}
\newcolumntype{P}[1]{>{\centering\arraybackslash}p{#1}}
\newcolumntype{M}[1]{>{\centering\arraybackslash}m{#1}}
\begin{document}

\title{Automated segmentation and extraction of posterior eye segment using OCT scans\\

\thanks{This work is supported by a research fund from Khalifa University. Ref: CIRA-2019-047 and the Abu Dhabi Department of Education and Knowledge (ADEK), Ref: AARE19-156. \textsuperscript{$\dagger$}Co-first Authors, \textsuperscript{*}Corresponding Author, Email: taimur.hassan@ku.ac.ae}
\vspace{-0.2cm}}
\author{\IEEEauthorblockN{Bilal Hassan\textsuperscript{1,$\dagger$}, Taimur Hassan\textsuperscript{2,$\dagger$,*}, Ramsha Ahmed\textsuperscript{3}, Shiyin Qin\textsuperscript{1,4}, and Naoufel Werghi\textsuperscript{2}},
\IEEEauthorblockA{\textsuperscript{1}School of Automation Science and Electrical Engineering, Beihang University (BUAA), Beijing 100191, China}
\IEEEauthorblockA{\textsuperscript{2}C2PS and KUCARS Centers in  Khalifa University of Science \& Technology, Abu Dhabi 127788, United Arab Emirates}
\IEEEauthorblockA{\textsuperscript{3}School of Computer and Communication Engineering, University of Science \& Technology Beijing, Beijing 100083, China}
\IEEEauthorblockA{\textsuperscript{4}School of Electrical Engineering and Intelligentization, Dongguan University of Technology, Dongguan 523808, China}
}

\maketitle
\begin{abstract}
This paper proposes an automated method for the segmentation and extraction of the posterior segment of the human eye, including the vitreous, retina, choroid, and sclera compartments, using multi-vendor optical coherence tomography (OCT) scans. The proposed method works in two phases. First extracts the retinal pigment epithelium (RPE) layer by applying the adaptive thresholding technique to identify the retina-choroid junction. Then, it exploits the structure tensor guided approach to extract the inner limiting membrane (ILM) and the choroidal stroma (CS) layers, locating the vitreous-retina and choroid-sclera junctions in the candidate OCT scan. Furthermore, these three junction boundaries are utilized to conduct posterior eye compartmentalization effectively for both healthy and disease eye OCT scans. The proposed framework is evaluated over 1000 OCT scans, where it obtained the mean intersection over union (IoU) and mean Dice similarity coefficient (DSC) scores of 0.874 and 0.930, respectively.
\end{abstract}
%%%\squeezeup
\section{Introduction}
Medical image analysis has been an essential tool in large number of organs pathology including   lung  \cite{taher_lung_12,taher_lung_13},  colon\cite{khatib_polyp_15,taha2017_surv, sajid2000},cervix \cite{taha_cervi_17}, and   prostate  \cite{reda_pros_16,alkadi_pros_19}. The eye, in particular, has been the subject of intensive research in medical imaging, especially with visual impairment being leading cause of disability, accounting for over 1 billion confirmed cases worldwide \cite{1hassan}. Out of many ocular disorders leading to the growing burden of vision loss, most disorders such as diabetic retinopathy, macular edema, and age-related macular degeneration are related to the posterior segment of the human eye \cite{2hassan2020rri}. As the name indicates, the posterior segment is the back part, covering the two-thirds region of the eye. It includes four compartments: vitreous, retina, choroid, and sclera \cite{12maloca2019validation}. Optical coherence tomography (OCT) is regarded as one of the most rapidly emerging imaging methods that allows adequate visualization of both anterior and posterior human eye segment \cite{12maloca2019validation,3Gholami,4Mahmudi}. However, the number of OCT scans acquired daily has surpassed the capacity of eye specialists in carefully examining these scans, necessitating automated methods to imitate the work of retinal experts \cite{5hassan2020seadnet}. 

In the past, many researchers have proposed automated algorithms to perform retinal image analysis using OCT scans. These works include detection and segmentation of chorioretinal features \cite{6TIM,7hassan2018automated}, classification and grading of retinal diseases \cite{8Kermany,9farisu,10hassan2019fully}, and retinal thickness measurement \cite{11cazanas2021ensemble}. However, to the best of our knowledge, very limited research has been conducted on the segmentation and extraction of posterior eye segment. One such work, inspired by the U-Net architecture, is proposed in \cite{12maloca2019validation}. While deep learning methods can achieve high accuracy they require larger data cohorts and extensive training time. Besides, the authors in \cite{12maloca2019validation} validated the feasibility of the work considering healthy eyes only. In contrast, this paper proposes an automated method built upon the standard image analysis techniques yet achieving comparable results for both healthy and disease eye OCT scans. The major contributions of this paper are as follows:
\begin{itemize}[leftmargin=*]
\item An efficient method for segmenting and extracting the posterior eye compartments (vitreous, retina, choroid, and sclera) for healthy and diseased eye cases using multi-vendor OCT scans.

\item An extensive evaluation using 1000 OCT scans, achieving state-of-the-art segmentation results with 0.874 mean intersection over union (IoU) and 0.930 mean Dice similarity coefficient (DSC) scores.
\end{itemize}

Based on the demonstrated segmentation and extraction results in this paper, the proposed framework can be of great value to upcoming OCT-based studies—for example, estimating retinal fluid in anti-vascular endothelial growth factor (anti-VEGF) therapy and measurement of retina and choroid thickness.

%%%%%%%%%%%%%%%%%%%%%%%%%%%%%%%%%%%%%%
\begin{figure*}[htb]
\centering
\includegraphics[width=1\linewidth]{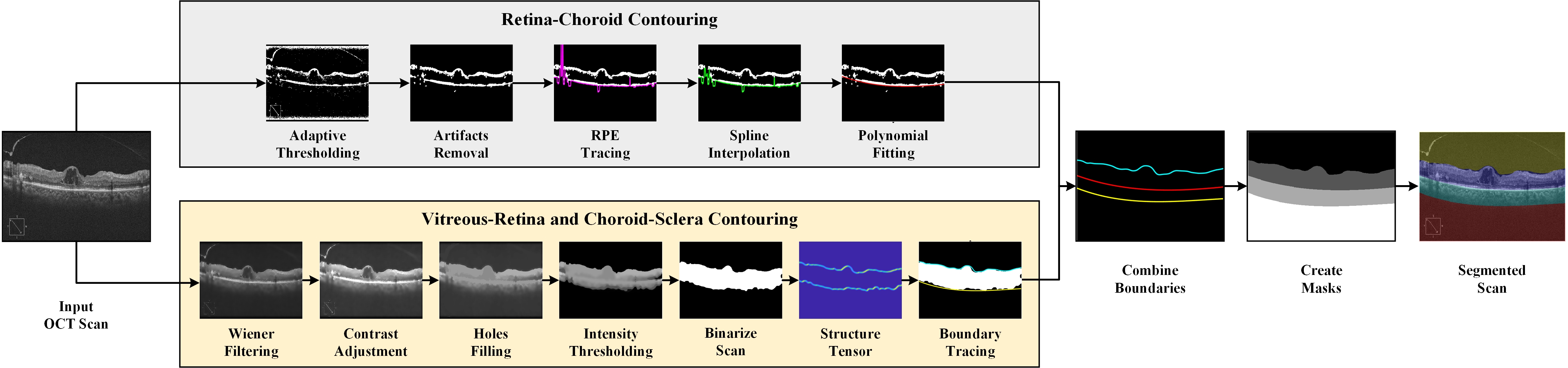}
\caption{Proposed framework for segmentation of posterior eye segment.}
\label{fig:fig1}
\end{figure*}
%%%%%%%%%%%%%%%%%%%%%%%%%%%%%%%%%%%%%%
%%%%%%%%%%%%%%%%%%%%%%%
\section{Proposed Methodology}\label{sec:model}
The raw OCT scans are pre-processed to convert them to grayscale and resize them to a common resolution of 360$\times$480 before input to the subsequent phases. The high-level overview of the proposed method is demonstrated in Fig. \ref{fig:fig1}, which consists of two main phases as described below:
\subsection{Retina-Choroid Contouring Phase}
We extract the RPE layer in this phase, which defines the retina-choroid junction in the candidate OCT scan. Since the RPE layer appears as a hyperreflective element in the OCT scan, the pre-processed input OCT scan is binarized using adaptive thresholding with ‘bright’ foreground polarity. It extracts the brighter contents from the candidate OCT scan, including the RPE layer, ILM layer, along with the unwanted speckle noise artifacts. These unwanted artifacts are removed from the scan using erosion and area opening operations in the next step. Next, to trace the RPE layer junction, a column-wise iterative approach is adopted to locate every last bright point in the OCT scan. Further, the zeros and ‘not a number’ values in the traced RPE layer vector are identified and removed using spline interpolation, as shown in Fig. \ref{fig:fig1}. In the final step, the smoothing operation is carried out using third-order polynomial fitting to obtain the retina-choroid junction.
\subsection{Vitreous-Retina and Choroid-Sclera Contouring Phase}
We extract the ILM and CS layers in this phase, defined as the respective vitreous-retina and choroid-sclera junctions in the candidate OCT scan. For this purpose, first, the pre-processed input OCT scan is denoised using Wiener filtering to suppress the speckle noise. Let $I$ be the pre-processed scan containing $N\times M$ sized local neighborhood ($\eta$) kernel for each pixel. Then Wiener filter estimates the local mean ($\mu$) and variance ($\sigma^2$) around each pixel as expressed in Eq. (1-3):
\begin{align}
F(n_1,n_2)=&\mu+\frac{\sigma^2+v^2}{\sigma^2}(I(n_1,n_2)-\mu), \\
\mu=&\frac{1}{NM}\sum_{n_1,n_2\in\eta}I(n_1,n_2), \\
\sigma^2=&\frac{1}{NM}\sum_{n_1,n_2\in\eta}I^2(n_1,n_2)-\mu^2,
\end{align}
where $F$ is the filtered scan and $v^2$ is the noise variance. Afterward, contrast adjustment is performed to enhance the chorioretinal region intensities followed by hole filling operation. Next, we perform intensity-wise thresholding where more weights are assigned to higher intensity values in the OCT scan to suppress the background and enhance the chorioretinal region, as shown in Fig. \ref{fig:fig1}. Then, it is binarized to apply the structure tensor ($S_T$) operation to determine the highly coherent tensor \cite{13hassan2019automated}. $S_T$ denotes the second-order moment matrix, which reflects image gradients with prevailing orientation using Gaussian derivative kernels within the given pixel neighborhood, as defined in Eq. (4):
\begin{equation}
S_T=\begin{bmatrix}
\omega*(\Delta P.\Delta P) & \omega*(\Delta P.\Delta Q)\\
\omega*(\Delta Q.\Delta P) & \omega*(\Delta Q.\Delta Q)
\end{bmatrix},
\end{equation}
where $\omega$ denotes the Gaussian smoothing filter, and $\Delta P$ and $\Delta Q$ represent the orientation of image gradients at zero and 90 degrees, respectively. Finally, Canny edge detection is applied to the highly coherent tensor to trace the ILM and CS layers in the final step, obtaining the respective vitreous-retina and choroid-sclera junctions.
\subsection{Segmentation of Posterior Eye Segment}
In the final phase, all the three boundary junctions (vitreous-retina, retina-choroid, and choroid-sclera, as shown in Fig. \ref{fig:fig1}) are combined and plotted on a zero-valued 360$\times$480 matrix gird. Finally, the posterior eye segment masks are created using different intensity values between each boundary junction to illustrate the segmentation results. Further, these segmentation results are overlaid on the pre-processed input OCT scan, as shown in Fig. \ref{fig:fig1}.
%%%%%%%%%%%%%%%%%%%%%%%
\section{Results}
\subsection{Dataset Details}
In this research, we have used 1000 OCT scans to evaluate the proposed framework. These scans are collected from four publicly accessible datasets (250 scans each). Farsiu et al. \cite{9farisu} curated the first dataset, which contains OCT scans obtained with the Bioptigen device. Gholami et al. \cite{3Gholami} arranged the second dataset, which consists of OCT scans imaged with the Cirrus machine. Similarly, Kermany et al. \cite{8Kermany} provided the third dataset with Spectralis OCT scans. The final dataset, organized by Mahmudi et al. \cite{4Mahmudi}, contains OCT scans of a Topcon device. The four datasets used in this study are named after the imaging devices they represent. Further, each OCT scan is manually labeled to generate the ground truth data. Table \ref{tab:1} summarizes the dataset details used in this study.
%%%%%% TABLE I %%%%%%%%%
\begin{table}[H]
\centering
\footnotesize
\caption{Dataset details}
\begin{threeparttable}
\setlength{\tabcolsep}{12pt}
\renewcommand{\arraystretch}{1.25}
\begin{tabular}{l|l}
\hline
Datasets	&	Pathologies (\textit{Scans}) 
	\\
\hline
Bioptigen \cite{9farisu}	&	Healthy (\textit{50}), AMD (\textit{200})	\\
Cirrus \cite{3Gholami}	&	Healthy (\textit{50}), CSR (\textit{75}), DR (\textit{75}), MH (\textit{50})	\\
Spectralis \cite{8Kermany}	&	Healthy (\textit{25}), AMD (\textit{75}), CNV (\textit{75}), ME (\textit{75})	\\
Topcon \cite{4Mahmudi}	&	Healthy (\textit{250})	\\
\hline
\end{tabular}
\begin{tablenotes}[flushleft]
\item\parbox[t]{8.6cm}{AMD = Age-related macular degeneration, CSR = Central serous retinopathy, DR = Diabetic retinopathy, MH = Macular hole, CNV = Choroidal neovascularization, ME = Macular edema.}
\end{tablenotes}
\end{threeparttable}
\label{tab:1}
\end{table}
%%%%%%%%%%%%%%%%%%%%%%%%%%
\subsection{Performance Metrics}
The performance of the proposed framework is assessed using four evaluation metrics, including accuracy, IoU, DSC, and boundary F1 (BF) scores. These are defined as follows:
\begin{align}
Accuracy=&\frac{TP+TN}{TP+TN+FP+FN}, \\
IoU=&\frac{TP}{TP+FP+FN}, \\
DSC=&\frac{2\times IoU}{1+IoU},\\
BF=&\frac{2\times TP}{2\times TP+FP+FN},
\end{align}
where $TP$, $TN$, $FP$, and $FN$ show the true positive, true negative, false positive, and false negative pixels, respectively.
\subsection{Qualitative Assessment}
\begin{figure}[htb]
\centering
\includegraphics[width=1\linewidth]{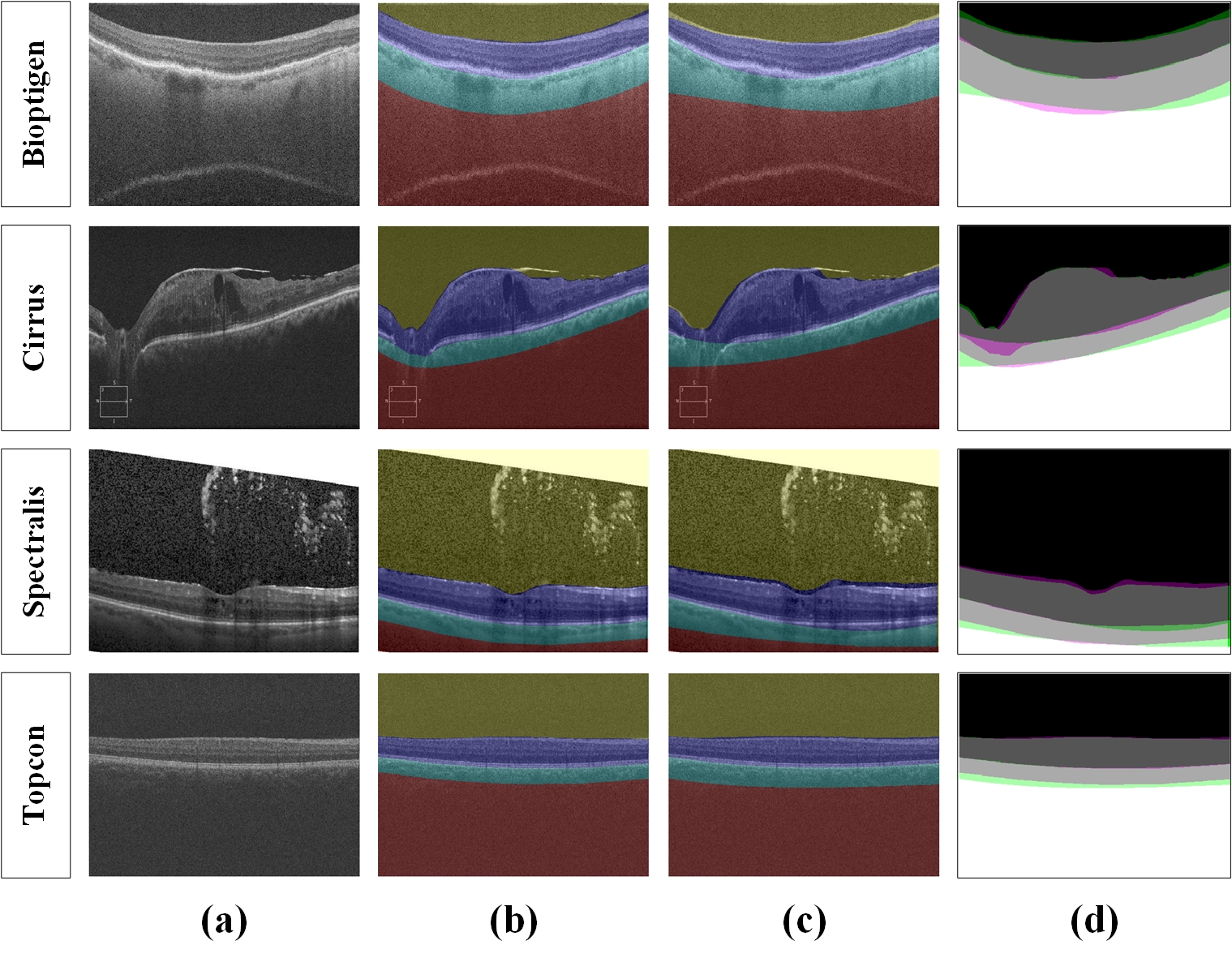}
\caption{Segmentation results. (a) Posterior segment OCT scans, (b) Ground truth labels, (c) Segmentation results, (d) Overlapped ground truth and segmented labels. Magenta shows false-positive pixels (over-segmentation), and green depicts the false-negative pixels (under-segmentation).}
\label{fig:fig2}
\end{figure}

This section presents the qualitative results for the segmentation of the posterior eye segment (vitreous, retina, choroid, sclera) by the proposed framework. In Fig. \ref{fig:fig2}, the segmentation results are showcased using randomly selected OCT scans from all four datasets investigated in this study. Here, we can observe that segmented regions are almost identical to the corresponding ground truth labels. It evidences the efficacy of the proposed method in producing excellent segmentation results regardless of the dataset and imaging devices. Further, it can be observed from Fig. \ref{fig:fig2}(d) that mainly the erroneous segmented pixels are in case of choroid class near the retina (RPE) and sclera (CS) junctions.  
%%%%%%%%%%%%%%%%%%%%%%%%%%%%%%%%%%%%%%
%%%%%%%%%%%%%%%%%%%%%%%%%%%%%%%%%%%%%%

\begin{figure}[htb]
\centering
\includegraphics[width=1\linewidth]{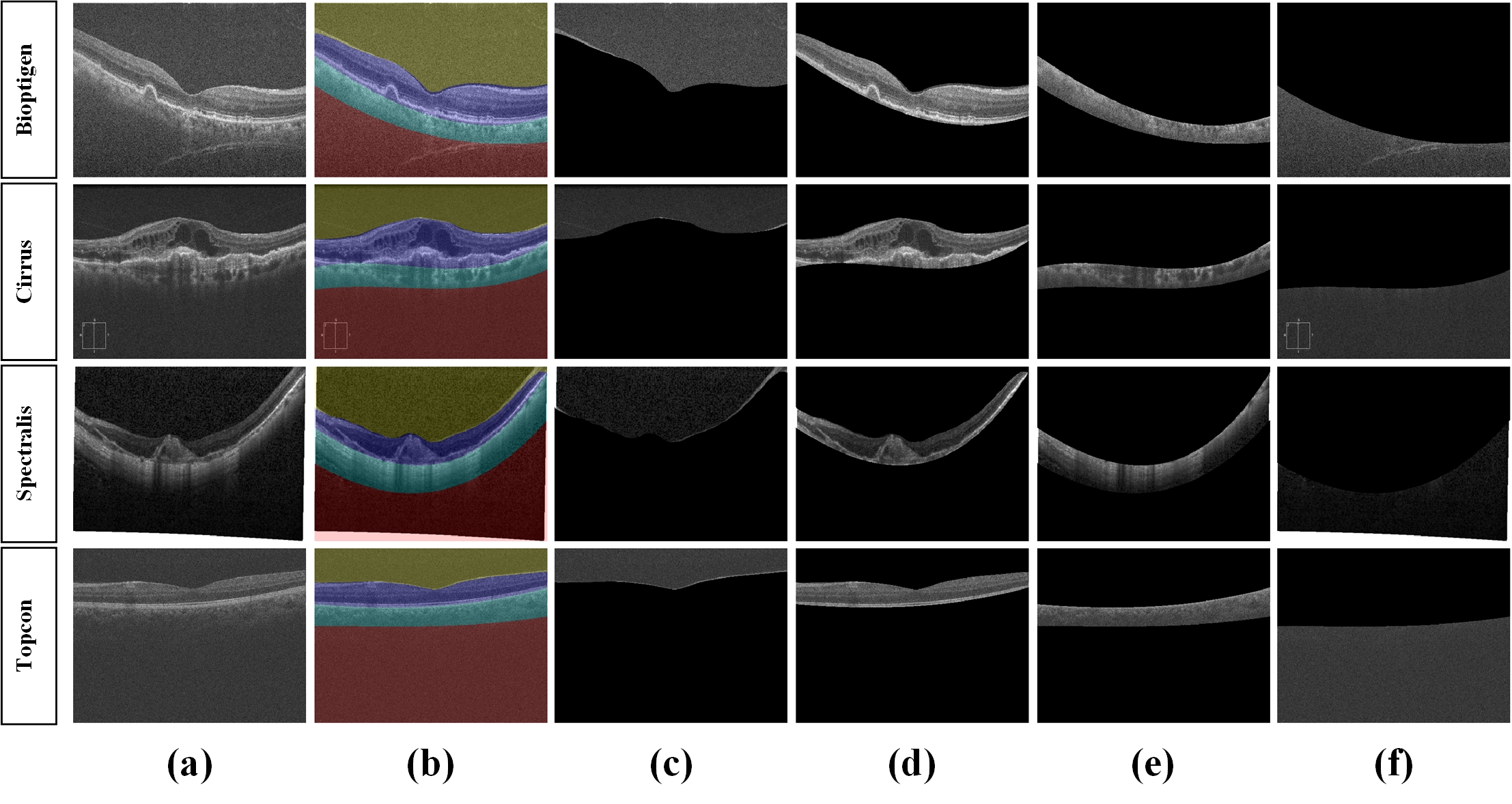}
\caption{Extraction results. (a) Posterior segment OCT scans, (b) Segmentation results. Extracted regions using segmentation masks (c) Vitreous, (d) Retina, (e) Choroid, (f) Sclera.}
\label{fig:fig3}
\end{figure}

Next, we present the extraction results of the posterior eye segment using the segmentation masks generated by the proposed method. The extraction of retina and choroid is crucial from the perspective of their thickness measurement and for better visualization of chorioretinal anomalies by discarding the artifacts in vitreous and sclera. The extraction results using randomly selected OCT scan from each dataset are shown in Fig. \ref{fig:fig3}. Here, we can appreciate the ability of the proposed method to precisely extract the posterior eye segment.
%%%%%%%%%%%%%%%%%%%%%%%%%%%%%%%%%%%%%%
%%%%%%%%%%%%%%%%%%%%%%%%%%%%%%%%%%%%%%
\subsection{Quantitative Assessment}
The performance of the proposed method is quantitatively assessed using various metrics. First, we measure the class-wise metrics to determine the proposed framework performance in terms of individual segmentation of compartments, as shown in Table \ref{tab:2}. Here, it can be observed that the best results are achieved for the vitreous class. In contrast, lower performance is observed for the choroid class.
%%%%%% TABLE II %%%%%%%%%
\begin{table}[htb]
\centering
\footnotesize
\caption{Class-wise metrics for segmentation of posterior eye segment. The best and the second-best results are shown in bold and underline, respectively.}
\begin{threeparttable}
\setlength{\tabcolsep}{10pt}
\renewcommand{\arraystretch}{1.25}
\begin{tabular}{l|c|c|c|c}
\hline
Metrics	&	Vitreous	&	Retina	&	Choroid	&	Sclera	\\
\hline
IoU	&	\textbf{0.969}	&	0.893	&	0.711	&	\underline{0.921}	\\
DSC	&	\textbf{0.984}	&	0.944	&	0.831	&	\underline{0.959}	\\
Accuracy	&	\textbf{0.987}	&	\underline{0.942}	&	0.936	&	0.925	\\
Mean BF Score	&	\textbf{0.902}	&	\underline{0.800}	&	0.539	&	0.701	\\
\hline
\end{tabular}
\end{threeparttable}
\label{tab:2}
\end{table}
%%%%%%%%%%%%%%%%%%%%%%%%%%

Next, we present the segmentation performance with respect to datasets and imaging devices, as shown in Table \ref{tab:3}. We can notice from Table \ref{tab:3} that the proposed framework achieved the best and second-best performance for the Topcon and Bioptigen OCT scans for all the metrics except weighted IoU, where the second-best value is obtained for Cirrus scans. On the other hand, it achieved a comparatively low performance for the Cirrus and Spectralis OCT scans. It is perhaps due to fewer variations in terms of retinal conditions (most of the scans belong to the healthy condition) in the Topcon \cite{4Mahmudi} and Bioptigen \cite{9farisu} datasets, making the segmentation task relatively easy for the proposed method. In contrast, the Cirrus \cite{3Gholami} and Spectralis \cite{8Kermany} datasets are more diversified, complex and contain OCT scans of different retinal pathologies (such as CSR, DR, MH, AMD, CNV, and ME); therefore, in this case, the segmentation of posterior eye segment is a much more complicated task and requires more accuracy. 
%%%%%% TABLE III %%%%%%%%%
\begin{table}[htb]
\centering
\footnotesize
\caption{Dataset-wise metrics for segmentation of posterior eye segment. The best and the second-best results are shown in bold and underline, respectively.}
\begin{threeparttable}
\setlength{\tabcolsep}{2.5pt}
\renewcommand{\arraystretch}{1.25}
\begin{tabular}{l|M{32pt}|M{30pt}|M{32pt}|M{32pt}|M{32pt}}
\hline
Metrics	&	Bioptigen \cite{9farisu}	&	Cirrus \cite{3Gholami}	&	Spectralis \cite{8Kermany}	&	Topcon \cite{4Mahmudi}	&	Overall	\\
\hline
Global Accuracy	&	\underline{0.946}	&	0.943	&	0.938	&	\textbf{0.968}	&	0.949	\\
Mean Accuracy	&	\underline{0.946}	&	0.943	&	0.925	&	\textbf{0.973}	&	0.948	\\
Mean IoU	&	\underline{0.882}	&	0.856	&	0.842	&	\textbf{0.909}	&	0.874	\\
Weighted IoU	&	0.902	&	\underline{0.903}	&	0.890	&	\textbf{0.943}	&	0.909	\\
Mean BF Score	&	\underline{0.747}	&	0.712	&	0.665	&	\textbf{0.817}	&	0.735	\\
\hline
\end{tabular}
\end{threeparttable}
\label{tab:3}
\end{table}
%%%%%%%%%%%%%%%%%%%%%%%%%%

\begin{figure}[htb]
\centering
\includegraphics[width=0.8\linewidth]{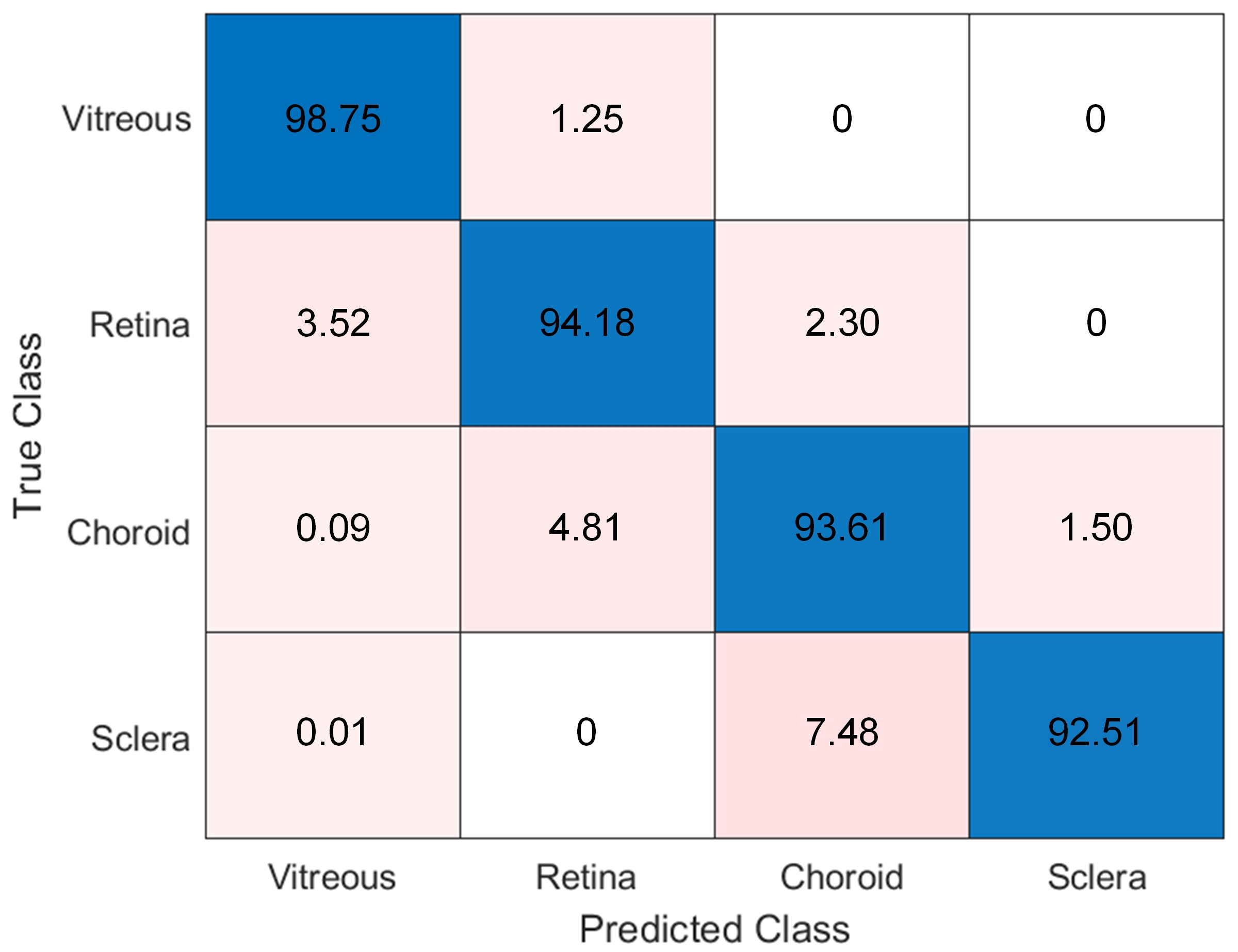}
\caption{Confusion matrix for pixel-level segmentation.}
\label{fig:fig4}
\end{figure}

Lastly, we compare the pixel-level results of the proposed framework with the corresponding ground truths using a normalized confusion matrix as shown in Fig. \ref{fig:fig4}. Here, the diagonal row showcases the percentage of correctly segmented pixels for each class. It can be seen from Fig. \ref{fig:fig4} that the proposed method faced the most confusion (around 8.98\%) between the sclera and choroid classes, where 7.48\% of the pixels in the sclera class are wrongly segmented as choroid pixels, and 1.50\% of the pixels in choroid class are incorrectly segmented as sclera pixels.  
%%%%%%%%%%%%%%%%%%%%%%%%%%%%%%%%%%%%%%
%%%%%%%%%%%%%%%%%%%%%%%%%%%%%%%%%%%%%%
\section{Conclusion}
This study proposed an automated method to segment the posterior eye compartments, including vitreous, retina, choroid, and sclera. Validated on 1000 scans from four different OCT imaging devices (Bioptigen, Cirrus, Spectralis, and Topcon), the proposed method demonstrated  effective performance in the posterior eye compartmentalization of retinal zones for both healthy and diseased eye candidate scans. Based on the demonstrated segmentation and extraction results, the proposed method adds substantially to the existing retinal image analysis literature. It offers exciting potential in terms of OCT scan compartmentalization for research purposes, observing treatment response, and tracking the disease progression in patients. In the future, this work can be extended to promote the method further to detect various chorioretinal irregularities, providing in-depth analysis of the underlying lesions in terms of detection, segmentation, and objective quantification.

\setstretch{1.26}
\footnotesize
\bibliographystyle{ieeetr}
\bibliography{refs}

\end{document}